\documentclass[epj,floatfix]{svjour}
\usepackage{amsmath}
\usepackage{amsmath}
\usepackage{amssymb}
\usepackage{amsfonts}
\usepackage{graphicx}
\usepackage{float}
\usepackage{wrapfig}
\usepackage{subfigure}
\usepackage{bm}
\usepackage{units}
\usepackage{grffile}
\usepackage{color}
\usepackage{hyperref}
\usepackage[title]{appendix}
\usepackage{academicons}
\usepackage{xcolor}

\begin{document}

\title{Taming Lagrangian Chaos with Multi-Objective Reinforcement Learning}

\author{Chiara Calascibetta\inst{1}\thanks{\href{calascibetta@roma2.infn.it}{email: calascibetta@roma2.infn.it (corresponding author)}}, 
Luca Biferale\inst{1}, 
Francesco Borra\inst{2}, 
Antonio Celani\inst{3}
\and Massimo Cencini\inst{4}}

\institute{
 Department of Physics \& INFN, University of Rome `Tor
Vergata', Via della Ricerca Scientifica 1, 00133 Rome, Italy \and
 Laboratory of Physics of the École Normale Supérieure, 24 Rue Lhomond, Paris, 75005, France\and
 Quantitative Life Sciences, The Abdus Salam International Centre for Theoretical Physics, ICTP, Trieste, 34151, Italy \and
 Istituto dei Sistemi Complessi, CNR, Via dei Taurini 19, Rome, 00185, Italy, and INFN `Tor Vergata'
}

\abstract{
  We consider the problem of two active particles in 2D complex flows with the multi-objective goals of  minimizing both the dispersion rate and the energy consumption of the pair. We approach the problem by means of Multi Objective Reinforcement Learning (MORL), combining scalarization techniques together with a Q-learning algorithm, for Lagrangian drifters that have variable swimming velocity. We show that MORL is able to find a set of trade-off solutions forming an optimal Pareto frontier. As a benchmark, we show that a set of heuristic strategies are dominated by the MORL solutions.  We consider the situation in which the agents cannot update their control variables continuously, but only after a discrete (decision) time, $\tau$. We show that there is a range of decision times, in between the Lyapunov time and the continuous updating limit, where Reinforcement Learning finds strategies that significantly improve over heuristics.  In particular, we discuss how large decision times require enhanced knowledge of the flow, whereas for smaller $\tau$ all \textit{a priori} heuristic strategies become Pareto optimal.
}

\titlerunning{Taming Lagrangian Chaos with Multi-Objective Reinforcement Learning}
\authorrunning{Calascibetta et al.}
\maketitle

\section{Introduction}\label{sec1}

In many engineering and geophysical applications robotic instruments
are often used for multi-agent sensing 
e.g. where a fleet of instrumented
drifters is used to collect information in the ocean,  multi-robots are used for searching sources leaking hazardous substances, or to probe complex
environments \cite{Lermusiaux_JMR,ELOR201276,WuWencen,Schmickl,Bechinger_2016}. A typical
application is how to keep the fleet under control, e.g., for patrolling the same region, keeping a given geometric formation and/or following a predetermined point-to-point  path. 
In typical flows, the relative distance between two passive
drifters would always grow, either  exponentially, due to Lagrangian chaos when
they are close, or in a diffusive way  at large scales, when non-linear effects become dominant \cite{crisanti1991lagrangian,cencini2010chaos}. Animal behaviour is often an inspiration and a leading direction of research trying to develop bio-mimetic strategies \cite{ginelli2016physics,Marchetti,Ballerini_birdflocks}. However, it is unclear whether using heuristic hard-wired rules  would be enough to control the swarm in the presence of a strongly  mixing flow \cite{khurana2013stability}. Moreover, in many realistic applications, agents need to take into account of strong engineering or biological limitations, needing to actively learn how to take advantage of the flow to accomplish the goal.  
As a result, we search to develop active complex policies to control complex environments. In chaotic or turbulent flows, the problem is given by the strong sensitivity of the system to any perturbation, making the very meaning of optimal control a fragile notion.
In this direction,  a few attempts to control single Lagrangian instrumented particles via Reinforcement Learning (RL) algorithms have been proposed to solve the Zermelo’s optimal navigation problem of reaching a fixed 
target \cite{Biferale_2019,Buzzicotti_Zermelo2020,bec2020}.
Moreover, RL has been successfully employed to
optimize the soaring of a glider in thermal currents 
\cite{Reddy_PNAS,reddy_nature} and to harness wind for airborne energy \cite{celani-airborne}. 
Recently, adversarial games between two competing agents have also been proposed to study chase-and-escape fish strategies at low Reynolds number \cite{Borra_2022}.

In this paper, we consider two agents (a particle pair) transported by a flow and having some limited knowledge on the underlying flow, which 
should act collectively so as to contrast the growth of their separation and, at the same time, to minimize as much as possible
the cost for control.
We assumed to solve the problem when the two particles  stay at distances
where the flow is differentiable so that, without control, their separation  would grow exponentially due to
Lagrangian chaos. To improve realism, we
model the problem imposing limitations in detection (partial observability) and in the possible actions to undertake (partial maneuverability). Namely, we allow each agent to sample only few local
properties of the underlying flow and to receive information about the other one
only at given \textit{decision times}, spaced by an interval $\tau$, in
order to update their actions. For what concerns the actions, following \cite{bec2020} we suppose that the two objects can swim
either along the direction of their separation or in the perpendicular
one, with a variable speed. Finally, to be able to fulfill the objective to minimize the energy cost, we also include the action of {\it no-swimming} to allow the couple of particles to learn when to be passively transported by the current, if useful \cite{Buzzicotti_Zermelo2020}.
As a result, we have a multi-agent (2 Lagrangian pair) and a multi-task (minimize both chaotic dispersion and energy consumption) problem \cite{Coello_1}. 
We approach this typical long-term optimization problem with conflicting objectives  by using Multi-Objective Reinforcement Learning (MORL) algorithms \cite{Coello_2,MORL_overview,Vamplew_MORL}.
 Indeed, in the classical single-task RL the reward is
a scalar, whereas in MORL the reward is a vector, with an element for each objective. We  approach MORL via scalarization, i.e. by defining a new scalar total reward by a  weighted sum
along all the element of the original reward vector
\cite{Natarajan_scalarization,Castelletti_scalarization}.
For this reason, there exists a set of trade-off solutions forming the so called Pareto frontier \cite{Vamplew_Pareto,Zitzler_Pareto}, where each solution on the frontier is Pareto Efficient, i.e. no single objective can be made better off without making at least another one worse off.

Using this approach, we show how to find a set of Pareto optimal policies to efficiently minimize both chaotic dispersion and swimming cost. To benchmark these strategies we will compare them to a set of heuristic baselines. In
particular, we show how the learned strategies are able to exploit nontrivial information of the underlying flow.

The paper is organized as follows. In Sec.\ref{sec2}, we describe the
general setup of the problem: the model of the Lagrangian pair, how
they can act and sense the environment, and the details of the
underlying fluid flow. In Sec.\ref{sec3} we introduce the concepts of
MORL giving details on our choice for the reward function and the
learning protocol. Furthermore, we present the concepts of Pareto
dominance and Pareto frontier. In Sec.\ref{sec4}, we discuss the main
results including a heuristic analysis focused on explaining the role
of the \textit{decision time}. Finally, we give our conclusion in
Sec.\ref{sec5}.

\section{The model}\label{sec2}

\subsection{Active control of Lagrangian pairs in a flow}\label{sec2.1}

In typical flows, Lagrangian chaos
\cite{crisanti1991lagrangian,cencini2010chaos} causes an exponential
growth of the separation, $\delta X(t)=\Vert \bm X_2(t) - \bm X_1(t)
\Vert$, between pairs of (uncontrolled) tracer particles that are
initially very close, i.e.  $\langle \ln\delta X(t)\rangle\simeq \ln
\delta (X(0)) + \lambda t$\,, ($\lambda$ being the Lagrangian Lyapunov
exponent).  Our goal here is to develop strategies for instrumented
particles to control and, possibly, minimize such chaotic dispersion
and at the same time to save energy. In particular, we consider
particles in the \text{one-way} coupling approximation and an
autonomous propelling mechanism with a speed $V(t)$ in the direction
$\hat{\bm p}(t)$ superimposed to the transport by the flow. Thus we
assume the particles to obey the following equations of motion:
\begin{equation}\label{eq:dynamics}
\begin{cases}
    \dot{\bm X_\alpha} = \bm u(\bm X_\alpha(t),t) + \bm U_\alpha^{ctrl}(t)\, ,  \\
    \bm U_\alpha^{ctrl}(t)=V_\alpha(t)\bm {\hat p}_\alpha(t),
\end{cases}\, 
\end{equation}
where $\alpha=1,2$ is the agent's index, $\bm u(\bm X_\alpha(t),t)$ is
the velocity of the underlying 2D advecting flow and $\bm
U_\alpha^{ctrl}(t)$ is the control contribution to the particle
velocity.

\begin{figure*}[!htb]
    \centering
    \includegraphics[width=0.7\textwidth]{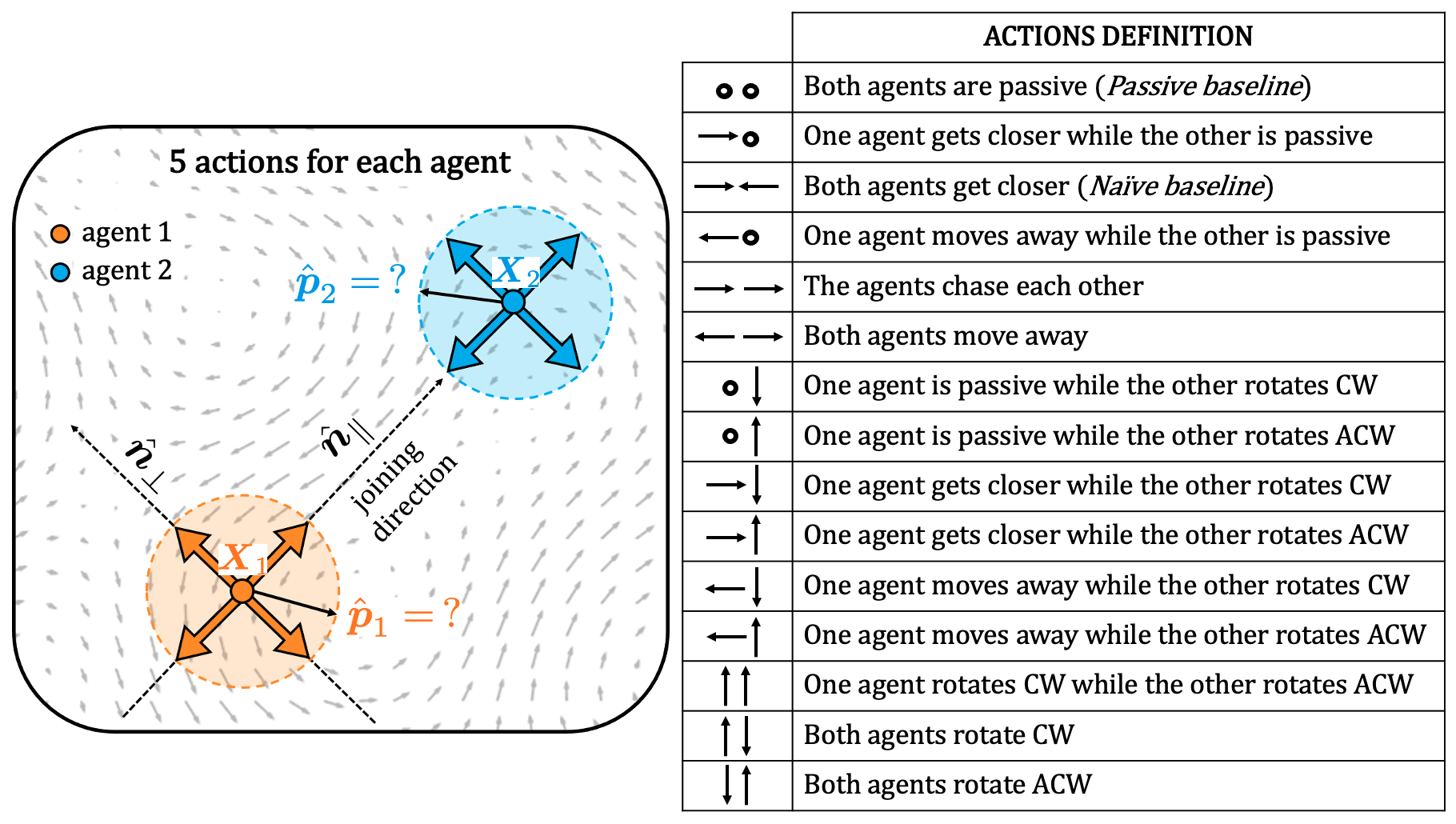} 
    \caption{(Left) The $5$ possible actions available to each agent,
      namely: remain passive or swim in the longitudinal direction
      $\pm\bm {\hat n}_\parallel$ or in the transversal one $\pm\bm
      {\hat n}_\perp$.  (Right) Scheme of the set of $15$ actions for
      the couple of agents, obtained after removing symmetrical action
      pairs. Besides the longitudinal actions, choosing the transversal directions allows the agents to rotate with respect to each other clock-wise (CW) or anti-clockwise (ACW).}
    \label{fig:Fig1}
\end{figure*}

We assume the agents to interact with the environment and between each
other only every $\tau$ time units. At each decision time the agents
can measure some flow properties (a proxy for the environmental
\textit{state}) and sense their mutual separation. Then, on the basis
of the information received, they can decide their \textit{action},
i.e. choose the swimming intensity and direction.  In this way, the
control $\bm U_\alpha^{ctrl}(t)$ becomes a piece-wise constant in time
function, i.e. $\hat{\bm p}_\alpha(t)=\hat{\bm p}_\alpha(t_j)$ and
$V_\alpha(t)=V_\alpha(t_j)$ for $t_j\leq t<t_j+\tau$, with $t_j=j\tau$
being the $j^{th}$ decision time.
Clearly, an
important role in achieving successfully strategies for staying close
is played by the dimensionless combination  of the two parameters $\tau \,\lambda$. When
$\tau\lambda>1$ the control is too sporadic and the velocity field can
separate considerably the agents. On the other hand, for $\tau\lambda
\to 0$,the control problem becomes easier. 

Concerning the swimming directions, $\hat{\bm p}_\alpha(t)$, we assume that
the agents have a limited set of choices. Namely, similarly to
\cite{bec2020}, the agents can either swim along their longitudinal
(joining) direction ($\hat{\bm n}_\parallel$) or in the transversal
one ($\hat{\bm n}_\perp$). Where $\hat{\bm n}_\parallel=({\bm X_2 - \bm
  X_1})/{\delta X}$
and $\hat{\bm n}_\perp \cdot \hat{\bm n}_\parallel = 0 $
(see Fig.~\ref{fig:Fig1}). We set the  swimming intensity, $V_\alpha(t)$  to be proportional to the agent distance (measured at the
decision time), i.e.
\begin{equation}\label{eq:velocity}
    V_\alpha(t)=F_\alpha(t)\delta X(t)\,,
\end{equation}
and we let each agent to choose either to turn on the control
by actively swimming, $F_\alpha(t)=f>0$, or to turn it off,
$F_\alpha(t)=0$, to save energy. 
With this choice, the first  objective is connected to  minimize the Lyapunov exponent of the controlled system. On the other hand, had we used a constant swimming speed, would have introduced a typical threshold distance above/below which the agents will always be able to control/not-control, at least, for small values of $\tau$.
Furthermore, to study the problem in a more challenging way,
we assume that swimming cannot completely overcome the dynamics, i.e. $2f\lesssim \lambda$ (see below).  

Summarizing, at each decision time $t_j$, agent $\alpha$ can pick any
of $5$ actions, $a\in \{F_\alpha(t_j)=0;F_\alpha(t_j)=f$
with $\hat{\bm p}_\alpha=\pm\hat{\bm n}_\parallel,\pm\hat{\bm
  n}_\perp\}$ namely, the agents can choose either to be passive or to
swim along their longitudinal or perpendicular directions.
We will call {\bf \textit{na\"{i}ve policy}} the strategy where the agents
always choose to navigate towards each other, i.e. $\hat{\bm
  p}_1=\hat{\bm n}_\parallel$ and $\hat{\bm p}_2=-\hat{\bm
  n}_\parallel$. Likewise, we will call {\bf \textit{passive policy}} the
strategy when  $F_1=F_2=0$.
In principle, a set of $5^2=25$ actions is available for the couple of
agents. However, the space of actions can be reduced by removing
symmetries (e.g. the configuration in which $\hat{\bm p}_1=\hat{\bm n}_\parallel$ and the second is
passive is equivalent to the configuration in which the first agent is
passive and  $\hat{\bm p}_2=-\hat{\bm
  n}_\parallel$). In this way, the set of actions for the couple
reduces to the set $\mathcal{A}$ of 15 actions shown in
Fig.\ref{fig:Fig1}.

\subsection{Sensing the environment}\label{sec2-sensing}
\begin{figure}[!htb]  
    \centering
\includegraphics[width=0.9\columnwidth]{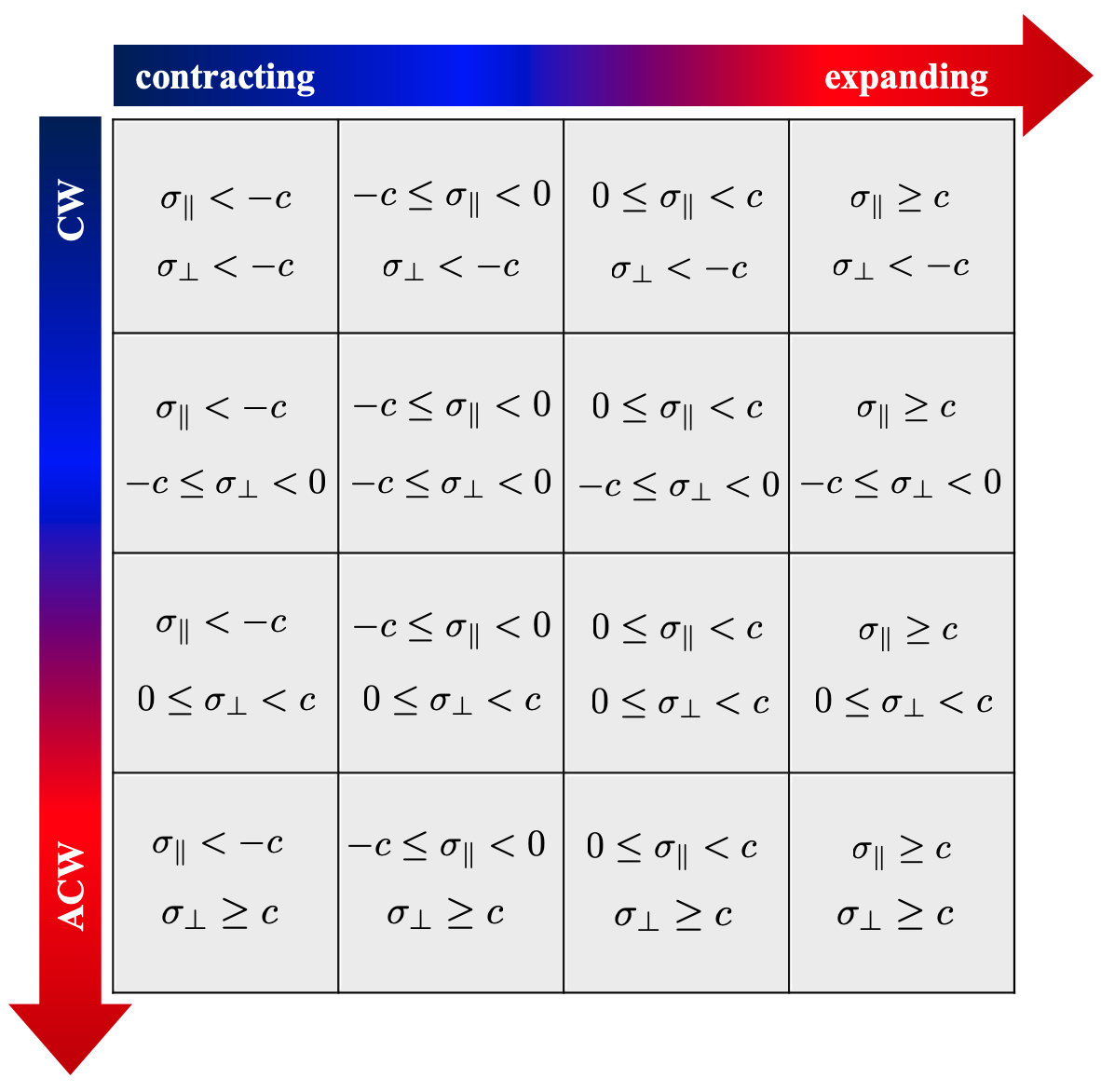} 
    \caption{Definition of the 16 states, obtained from the
      discretization of longitudinal (\ref{eq:long_vel}) and
      transversal (\ref{eq:perp_vel}) gradients. The former informs
      about the local rate of expansion/contraction due to the flow,
      the latter tells the rate of clock-wise (CW) or anti-clockwise
      (ACW) rotation imposed by the flow on the joining direction.  In
      figure is shown the states discretization, with $c$ a
      discretization constant.}
    \label{fig:Fig2}
\end{figure}

Besides their relative position and distance, at each decision time,
the agents receive some cues on the fluid environmental state. 
Concerning the observability of the environment,
assuming that they are close enough for the field to be smooth, we imagine the two agents can have only a rough estimates of the relative longitudinal and transverse gradients, defined as
\begin{equation}\label{eq:long_vel}
    \sigma _\parallel = \frac{(\bm u(\bm X_2,t)-\bm u(\bm X_1,t))}{\delta X}\cdot \hat{\bm n}_\parallel\,
\end{equation}
\begin{equation}\label{eq:perp_vel}
    \sigma_\perp= \frac{(\bm u(\bm X_2,t)-\bm u(\bm X_1,t))}{\delta X}\cdot \hat{\bm n}_\perp\,,
\end{equation}
which can be obtained by exchanging information about their local velocities.
To further  restrict the state-space, we suppose that the agents are able to measure their velocity difference and separation, as gradients approximation, with a limited sensitivity and  we restrict the set of values of $\sigma_\perp,\sigma _\parallel$ to 4 states for each one of them, for a total of 16 discretized states, labeling whether the underlying flow brings them closer or farther away (longitudinal gradients) and in which direction it rotates them (transverse gradients)
see Fig.~\ref{fig:Fig2} for a summary of all states. 
The value of the discretization constant $c$ is
chosen such as all the $16$ states are sufficiently visited ($c=2.8$ in our case).

\subsection{The space of control policies}\label{sec2-policies}

Given 15 actions and 16 states we have a possible set of $15^{16}$ deterministic policies $[\pi : s \to a]$, making the brute force optimization search impossible: one has to
resort to Reinforcement Learning techniques (as discussed in Sec. \ref{sec3}). However, there is an intuitive way to trace back to a reduced policies space that can be analyzed systematically as a set of hardwired baseline policies. By restricting perceptions to the longitudinal components  (\ref{eq:long_vel}) we can assume that only the actions along the joining direction are important. Considering only the first three actions in the right panel in Fig.\ref{fig:Fig1}, we get to a $3^4 = 81$ reduced set. In the following, we will consider these $81$ policies as our reference heuristics to benchmark the one found by our MORL implementation.

\subsection{Model flow}\label{sec2.2}
As for the fluid environment, we used a 2D homogeneous, (nearly)
isotropic, incompressible and time-dependent flow as in
Ref.~\cite{bec2005}. In particular, the velocity field is defined in
terms of a stream function, $\bm u(\bm x,t)=\nabla_\perp\psi(\bm x,t)
= (\partial_y \psi,-\partial_x \psi)\,,$ which is expressed as a
superposition of few Fourier modes,
\begin{equation}\label{eq:streamfunction}
\psi(\bm x,t)=\sum_{\bm k \in {\mathcal{K}}}{(A(\bm k,t)e^{i\,\bm k\cdot \bm x}+cc.)\,,}
\end{equation}
${\mathcal{K}}=\{(\frac{2\pi}{L},0),(\pm\,\frac{2\pi}{L},\frac{2\pi}{L}),(0,\frac{2\pi}{L})\}$, where $L$ is the scale periodicity of the flow.
In \eqref{eq:streamfunction}  $A(\bm k,t)=A_r(\bm k,t)+i\,A_i(\bm k,t)$ are random and time-dependent amplitudes obtained from an Ornstein-Uhlenbeck process \cite{OUprocess} 
\begin{equation}\label{eq:Amplitudes}
    \dot{A}_\beta(\bm k,t)=-\frac{1}{\tau_f}A_\beta(\bm k,t)+\Big{(}\frac{2\sigma^2(\bm k)}{\tau_f}\Big{)}^{1/2}\eta_\beta(\bm k,t)\,,
\end{equation}
with $\beta=r,i$. Where $\tau_f$ sets the flow correlation time,
$\eta_\beta(\bm k,t)$ are zero-mean Gaussian variables with
correlation $\langle \eta_\alpha(\bm k,t)\eta_\beta(\bm
k^\prime,t^\prime)\rangle=\delta_{\alpha,\beta}\delta_{\bm k,\bm
  k^\prime}\delta(t-t^\prime)$\,, and $\sigma^2(\bm
k)=\frac{u^2_{rms}}{2\Vert \bm k\Vert^2}$. We fixed
$\tau_f=1,L=1,u_{rms}=1$. With this choice the maximum Lyapunov
exponent characterizing the mean exponential rate of divergence
between two (uncontrolled) tracers particles is $\lambda \simeq 1.4$.

\section{Reinforcement learning and multi-objective optimization}\label{sec3}
\begin{figure}[!t]
    \centering
\includegraphics[width=0.5\textwidth]{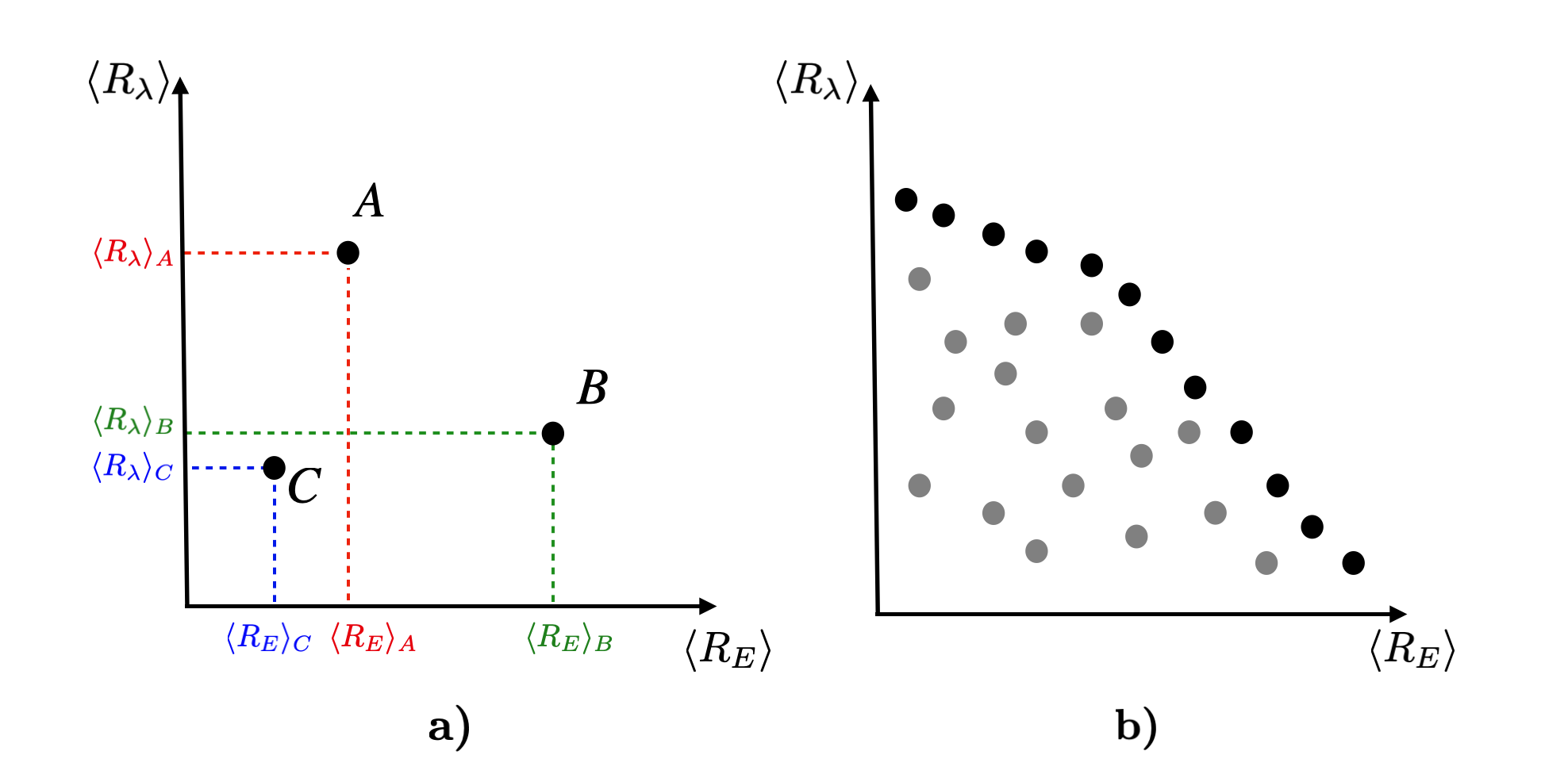}\caption{Concepts of Pareto dominance \textbf{(a)} and Pareto frontier \textbf{(b)}. $\langle R_\lambda\rangle$ and $\langle R_E\rangle $ represent the two single objectives of minimizing particle separation and power consumption. }
    \label{fig:Fig3}
\end{figure}

Starting from our set of states and actions, our aim is to solve an
optimization problem with two objectives: minimizing both the rate of
separation growth and the energy consumption. We are thus in the field
of competing Multi-Objective Optimization (MOO) \cite{Coello_1,Coello_2}.
The optimality of such solutions can be
defined in terms of Pareto dominance \cite{Vamplew_MORL,Zitzler_Pareto}, namely a
solution dominates another if it is superior on at least one objective
and at least equal on all others. For instance, in Fig.\ref{fig:Fig3}
a) the solutions $A$ and $B$ dominate $C$, whereas $A$ and $B$ are
incomparable, because each is superior in at least one objective. All
the dominating solutions form the Pareto frontier
\cite{Vamplew_MORL,Zitzler_Pareto}, depicted with black circles in
Fig.\ref{fig:Fig3} b).

Reinforcement Learning (RL) algorithms \cite{sutton2018reinforcement}
aim at maximizing a single scalar reward  usually representing a single long-term objective. 
MOO can also be obtained within standard RL
algorithms such as, e.g., Q-learning \cite{sutton2018reinforcement} 
by scalarization,  formulating a ``new'' total single-objective optimization problem obtained  as a
weighted sum of each sub-objective functions \cite{Natarajan_scalarization,Castelletti_scalarization}.
By solving the scalar optimization problem at varying the weights in the sum one can find the Pareto optimal solutions to the MOO.
Following this idea, we
define a different reward function for each of the two competing sub-problems.

The first allows the agents to judge their performance in controlling
the separation rate:
\begin{equation}\label{eq:rlambda}
r_\lambda(t_j)=-\frac{1}{T_{max}}\ln\Big{(}\frac{\delta X(t_j)}{\delta X(t_j-\tau)}\Big{)}\,,
\end{equation}
which penalizes actions that, between two consecutive decisions, cause an increase of the distance, and where $T_{max}$ is a fixed time horizon
that we considered as terminal state for the learning episodes and chosen such that the relative distance between the two particles is always in the linear regime ($T_{max}=5.5$ in our case). 
Notice that summing the reward \eqref{eq:rlambda} over a whole episode
\begin{equation}
    R_\lambda=\sum_{j=1}^{T_{max}/\tau}{r_\lambda(t_j)}
    \label{eq:Rtot}
\end{equation} 
when averaging over may episodes we 
have
\begin{equation}\label{eq:totalRlambda}
    \langle R_\lambda \rangle = - \frac{1}{T_{max}}\Big{\langle} \frac{\delta X(T_{max})}{\delta X(0)} \Big{\rangle} \simeq -\lambda_c\,.
\end{equation}
Thus, the optimization problem restricted to this reward would be solved by the policy which minimizes the Lyapunov exponent of the controlled system, $\lambda_c$.

The second reward function informs the agent about the energy cost:
\begin{equation}\label{eq:rE}
r_E(t_j)=-\frac{1}{T_{max}}\lambda\tau N_a(t_j)\,
\end{equation}
where $N_a(t_j)(=0,1,2)$ counts the number of agents 
which have selected any of the actions `swim'; we have  introduced a normalization term $\lambda\tau$ to have two rewards of the same order of magnitude (on average we can estimate $r_\lambda=(\tau \lambda)/T_{max}$).

For the multi objective optimization,  we need to combine \eqref{eq:rlambda} and
\eqref{eq:rE} through a scalarization parameter, $\beta$:
\begin{equation}\label{eq:rtot}
r_{tot}(t_j)=r_\lambda{(t_j)}+\beta r_E(t_j)\,
\end{equation}
and  consider many single-objective problems for $r_{tot}$ at varying $\beta$. Therefore, at each
decision time, the Lagrangian pair receives a shared reward,
$r_{tot}$.
For each $\beta$ the goal is to find the policy  maximizing the cumulative total reward, 
\begin{equation}
R_{tot} = \sum_{j=1}^{T_{max}/\tau} [r_\lambda(t_j)+\beta\,r_E(t_j)]=R_\lambda+\beta R_E  \label{eq:totreward}\,.
\end{equation}
From the above expression there are two clear limits:
$\beta\rightarrow0$ and $\beta\rightarrow\infty$. In the former limit we  minimize particle distance without caring on energy consumption, a goal that is not obvious {\it per se} and will depend on the decision time, $\tau$.  The latter case is simpler, because as the cost of
swimming increases the best policy is the passive one. How does the transition between these two limiting regimes take place is the question that we are going to
answer in Sec.\ref{sec4} by studying the Pareto frontier of our multi-task problem. See Appendix \ref{appendixA} for details of the Q-learning algorithm that we have implemented. 

Due to learning stochasticity and the fact that the performances of different policies can be very close,  to find the best solutions  we performed $100$ independent learning  sections  (i.e. using different initialization seeds) for each
scalarization parameter $\beta$. Each learning section lasts  $50000$ episodes. The 100 learned policies for each $\beta$ have been validated on  100000
different realizations of the flow.    Then we assumed as \textit{best
  learned policies} the ones that maximize the average over the validation set of 
(\ref{eq:totreward}).

\section{Results}\label{sec4}
Here, we present the results. In the first section, we present a
detailed analysis of the policy related to a fixed decision time
value, $\tau=0.3$, which given that the uncontrolled Lyapunov exponent
is $\lambda\approx 1.4$ corresponds to a case $\tau\lambda<1$ (and
$\tau<\tau_f=1$) that allows us to control the dynamics but the
interval between decision times, $\tau$, is sufficiently high to make the problem nontrivial.  In
the second section, we analyze specifically the role of $\tau$ presenting a heuristic analytical prediction in the $\tau\to 0$ limit and a numerical study based on the heuristic
policies as a function of $\tau$. In particular, we show that  
the concavity of the Pareto frontier is strongly dependent on $\tau$. 

To make the control non-trivial, we fix the swimming rate to $f=0.7$ so
that the controlled Lyapunov exponent is never negative. We also
tested smaller values obtaining qualitatively similar results.

\subsection{Detailed analysis of learned policies}\label{sec4.1}
\begin{figure}[!t]
    \centering
\includegraphics[width=0.5\textwidth]{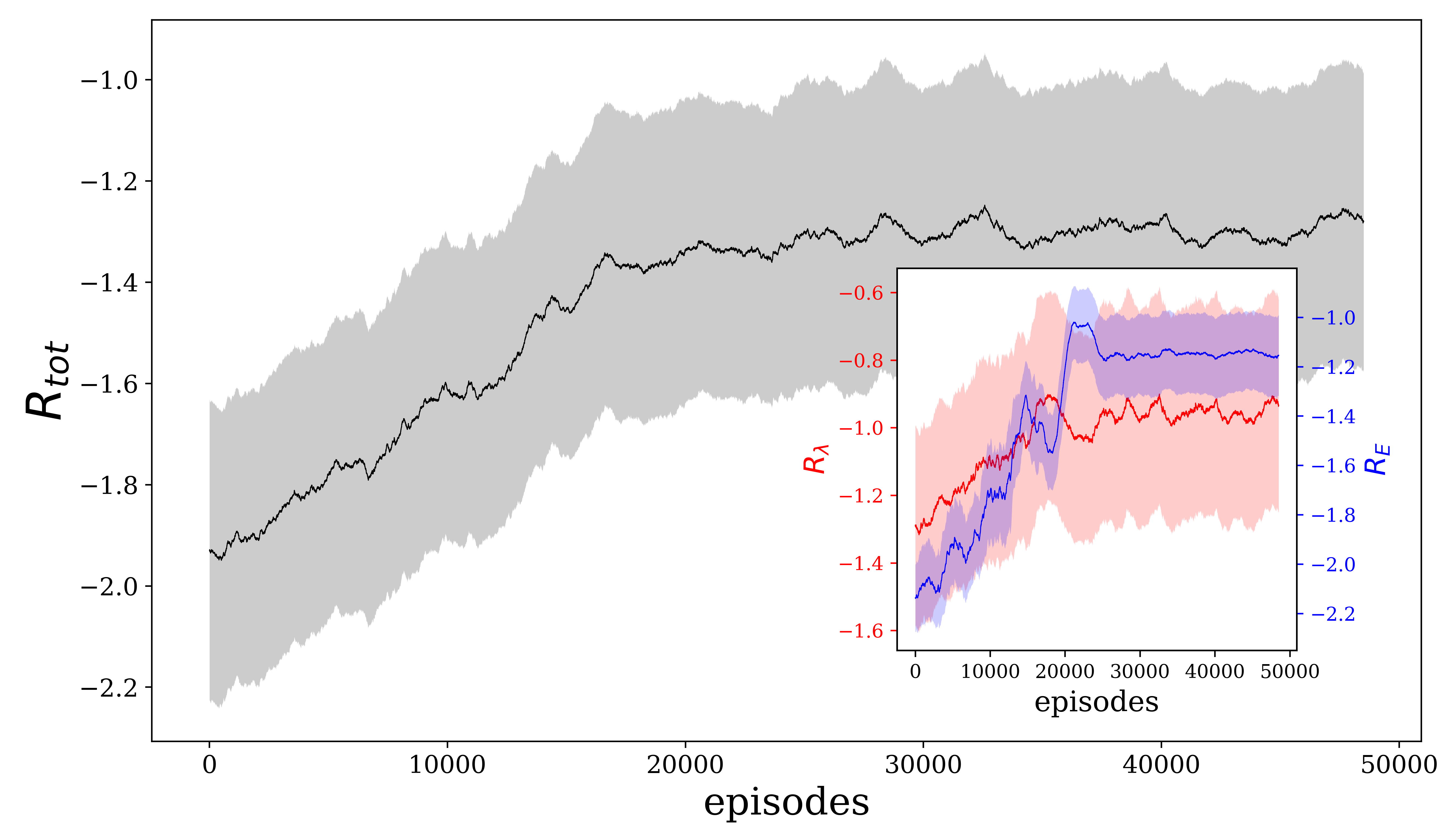}\caption{Example of
  learning process. The total reward $R_{tot}$ vs the episodes: the
  solid line show the running average over 1500 episodes while the
  gray shaded area shows the fluctuations in the individual
  episodes. The inset show the evolution of the two contributions to
  the reward $R_\lambda$ (red) and $R_E$ (blue). Data refer to one
  learning trial for $\beta=0.3$ and $\tau=0.3$.}
    \label{fig:Fig4}
\end{figure}

As discussed in Sec.~\ref{sec3}, we approach the multi objective
optimization through reinforcement learning via scalarization, that is
we perform many learning processes, with the protocol described at the
end of Sec.~\ref{sec3}, by varying the scalarization parameter $\beta$
that weighs the two rewards (see \eqref{eq:totreward}).  In
Fig.~\ref{fig:Fig4}, we show a typical learning process for a single
$\beta$ value.  At the beginning of the learning phase, the Q-learning
algorithm explores different random policies. After many episodes the
learning parameters, $\epsilon$ and $\alpha$, decrease (see
\eqref{eq:alpha}-\eqref{eq:eps}) and the Q-matrix stabilizes. The inset of 
Fig.~\ref{fig:Fig4} displays the evolution of two components of the reward.
As discussed, for the same $\beta$ the learning process is then repeated over
$100$ trials, each learned policy is evaluated on a validation set to identify
the best learned policy for that value of $\beta$.

\begin{figure*}[!bht]
    \centering
    \includegraphics[width=0.85\textwidth]{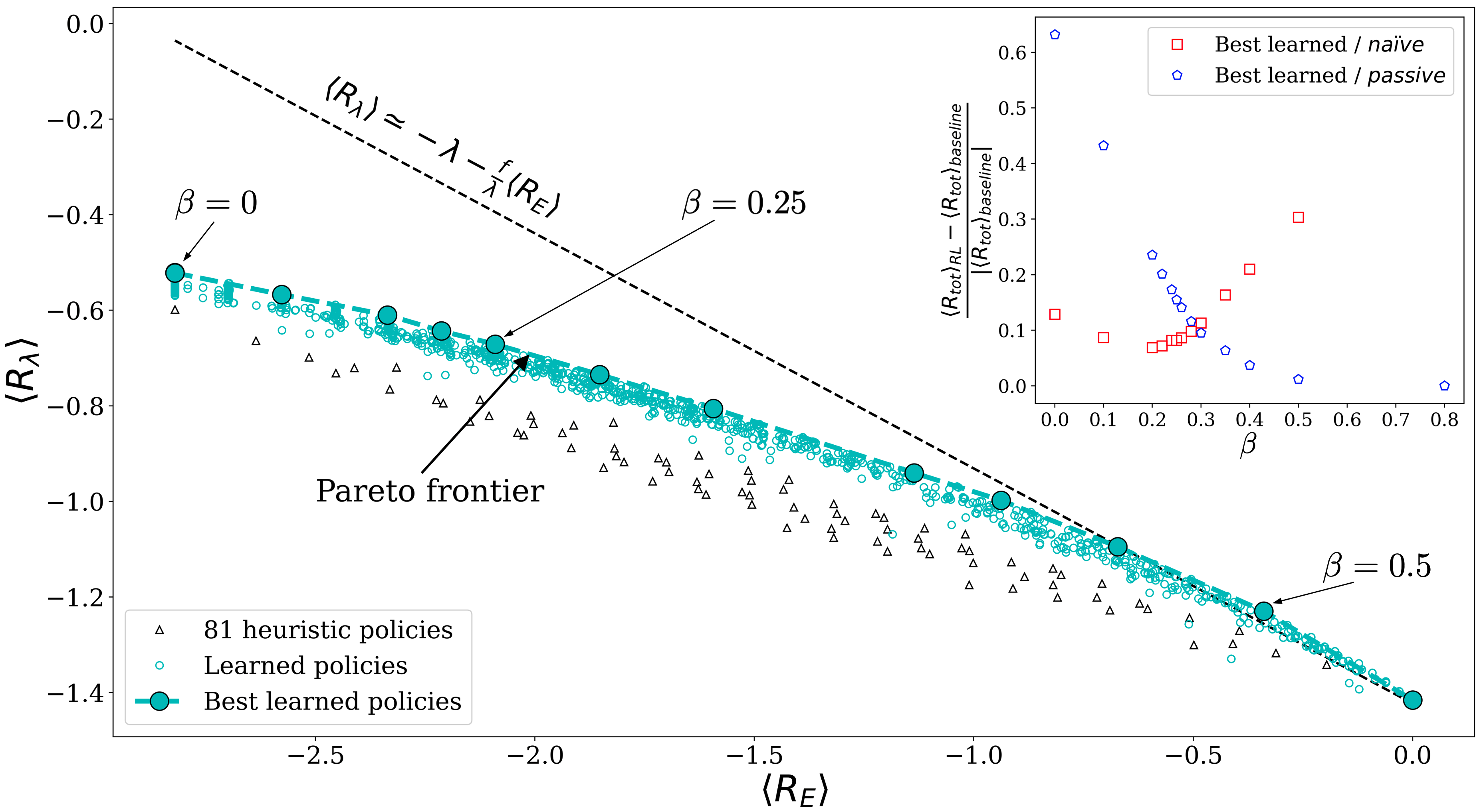} 
    \caption{$\langle R_\lambda \rangle$ vs $\langle R_E \rangle$
      obtained by setting the decision time $\tau=0.3$. The dashed
      black line indicates the highest performances that can be
      achieved for $\tau\rightarrow0$ (see \eqref{eq:linearity_Rlambda_Re}). We show with black triangles the performances of 81
      heuristic policies and with light blue circles all the policies
      learned from the Q-learning algorithm to solve the
      multi-objective problem, in particular the filled large circles show
      the best learned policies for each $\beta$, and the dashed blue
      line the Pareto frontier. One can see that learned policies
      dominate the heuristic ones. Inset: Relative improvement of
      $\langle R_{tot}\rangle$ as a function of $\beta$ of the best
      policies learned from RL with respect to the \textit{na\"{i}ve}
      (red squares) and passive (blue pentagon), respectively. }
    \label{fig:Fig5}
\end{figure*}

In Fig.\ref{fig:Fig5}, we show the mean (over the validation set) of
the total cumulative reward obtained to minimize the separation,
$\langle R_\lambda \rangle$ \eqref{eq:totalRlambda} vs the mean of the
total cumulative reward obtained to save energy, $\langle
R_E\rangle=\Big{\langle}\sum_{j=1}^{T_{max}/\tau}{r_E(t_j)}\Big{\rangle}$,
for different $\beta$ values. Then we can identify as optimal the
solutions that dominate all the others in the sense of Pareto
dominance. In particular the best learned policies for each $\beta$ are
surely on the Pareto frontier (filled circles in
Fig.~\ref{fig:Fig5}). As one can see, the learned policies reach
rewards that outperform the 81 heuristic policies, indicating that the
RL protocol has converged to optimal solutions. The inset in
Fig.\ref{fig:Fig5} provides a quantitative idea of the improvement with
respect to the \textit{na\"{i}ve} and passive baseline,
respectively, by showing the improvement of the
best-learned policies as a function of $\beta$. One can see that the
best learned policies are generically better than the baselines.
\begin{figure*}[!htb]
\centering
\includegraphics[width=0.75\textwidth]{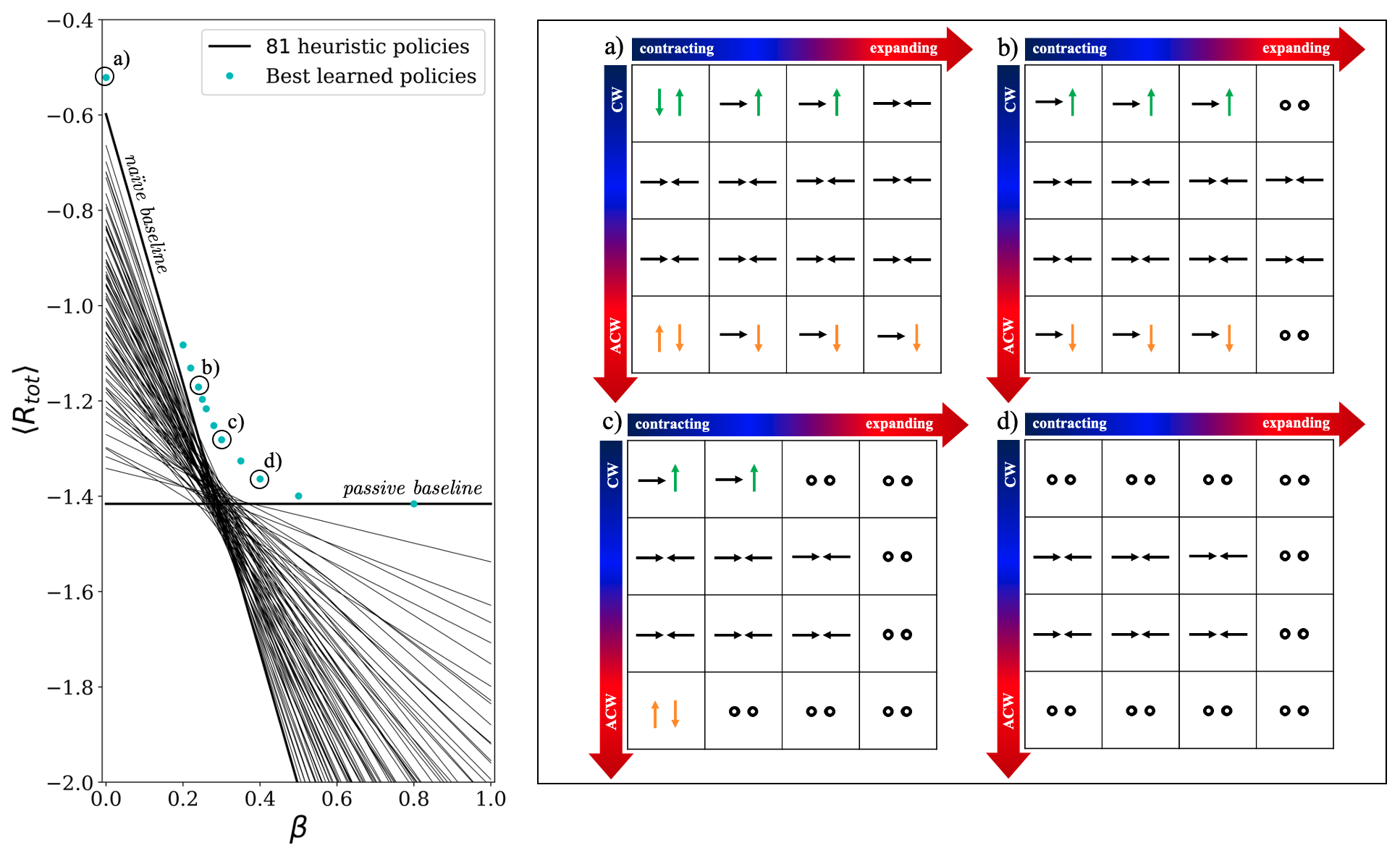} 
\caption{(Left) $\langle R_{tot} \rangle$ vs $\beta$ for different
  policies obtained with a decision time $\tau=0.3$. With the 81
  straight lines are shown the heuristic performances, including the
  two extreme cases of the \textit{na\"{i}ve} baseline and the passive
  policy.  It can be seen that as $\beta$ increases the
  \textit{na\"{i}ve} policy immediately get worse, while the passive
  policy is $\beta$-independent. In light blue are shown the best
  policies learned from Q-learning for each $\beta$ that improve all
  the heuristic performances. (Right) Pictorial representation of $4$
  policies learned from RL corresponding to the circled symbols on the
  left. In green and orange are highlighted the ACW and CW rotation
  actions chosen as a counter-move with respect to the underlying flow
  rotation.  }
    \label{fig:Fig6}
\end{figure*}
In particular, for $\beta=0$, since there is no cost for swimming, one
might expect the \textit{na\"{i}ve} baseline to be a good strategy for
minimizing the agents separation. Instead, we discovered that also in this limit
 there is a nontrivial optimal strategy for Lagrangian agents
that outperform the \textit{na\"{i}ve} one with an improvement of $12\%$
in the maximization of the total reward. The discovered strategy is such that, when the
velocity field is contracting and rotates strongly the position of one
agent with respect to the other, it turns out to be more convenient to
counter-rotate with respect to the rotation induced by underlying velocity field 
rather than to simply navigate towards each other. We can explain that as a
realignment along the contracting direction. Indeed, due to the finite
decision time, it is less effective to swim towards each other in the
direction identified at the decision time, which is quickly changed by
the flow.  On the other hand, for high values of $\beta$, when
swimming has a high cost, RL easily learns that the
best policy is the passive one, where the two agents always switch off
the engine.  Other $\beta$ values lead to the transition region
between these two extreme cases, where new navigation strategies are
learned. In particular, as seen from the inset of Fig.~\ref{fig:Fig5}, the region around $\beta=0.3$ seems
to be the more interesting, it is thus worth analyzing the learned
policies in this region.

In Fig.\ref{fig:Fig6}(Left) we show the performance of the best
learned policies for each $\beta$ in comparison with the 81 heuristic
ones. Notice that for the latter it is enough to measure $\langle
R_\lambda\rangle$ and $\langle R_E\rangle$ to know the value of
$\langle R_{tot}\rangle$ as a function of $\beta$, obtaining 81
benchmark straight lines, as shown in Fig.\ref{fig:Fig6}(Left). It can
be seen that RL always improve the maximization of the total
cumulative reward, $\langle R_{tot}\rangle$.  In
Fig.\ref{fig:Fig6}(Right)a-d we show a tabular representation of some
policies learned as optimal, i.e. that lie on the Pareto frontier and
appertain to the region around $\beta=0.3$ (see circled dots in
Fig.\ref{fig:Fig6}(Left)). It emerges that counter-rotating with
respect to the underlying flow rotation is important and that, as
$\beta$ increases, is more convenient navigate when the flow is
contracting rather than when it is expanding the agents' separation;
e.g. the policy in Fig.\ref{fig:Fig6}c) shows that the Lagrangian pair
choose to be passive along all the expanding states.

We end this section  commenting on possible symmetries  of swimming
strategies.  In fact, the counter-rotating action with respect to the
underlying flow rotation that emerges to be important can occur both
when the counter-rotation is clockwise and when it is
anti-clockwise. This could lead to some ambiguity in the choice of swimming
strategies or, more precisely, it could happen that equivalent
strategies exist simultaneously. However, this does not seem to cause
convergence issues, and RL is still able to identify optimal
solutions.

\subsection{Heuristic analysis}\label{sec4.2}
We now discuss the role of the interval between decision times, $\tau$, by relying on the
set of 81 heuristic (hard-wired) strategies obtained as a reduced
policies space from the 4 longitudinal states and the first 3 actions
in Fig.\ref{fig:Fig1}. Indeed, their analysis is enough to understand
the qualitative effect of changing the decision time with no needed to
perform any learning.
\begin{figure}[!t]
\centering
\includegraphics[width=\columnwidth]{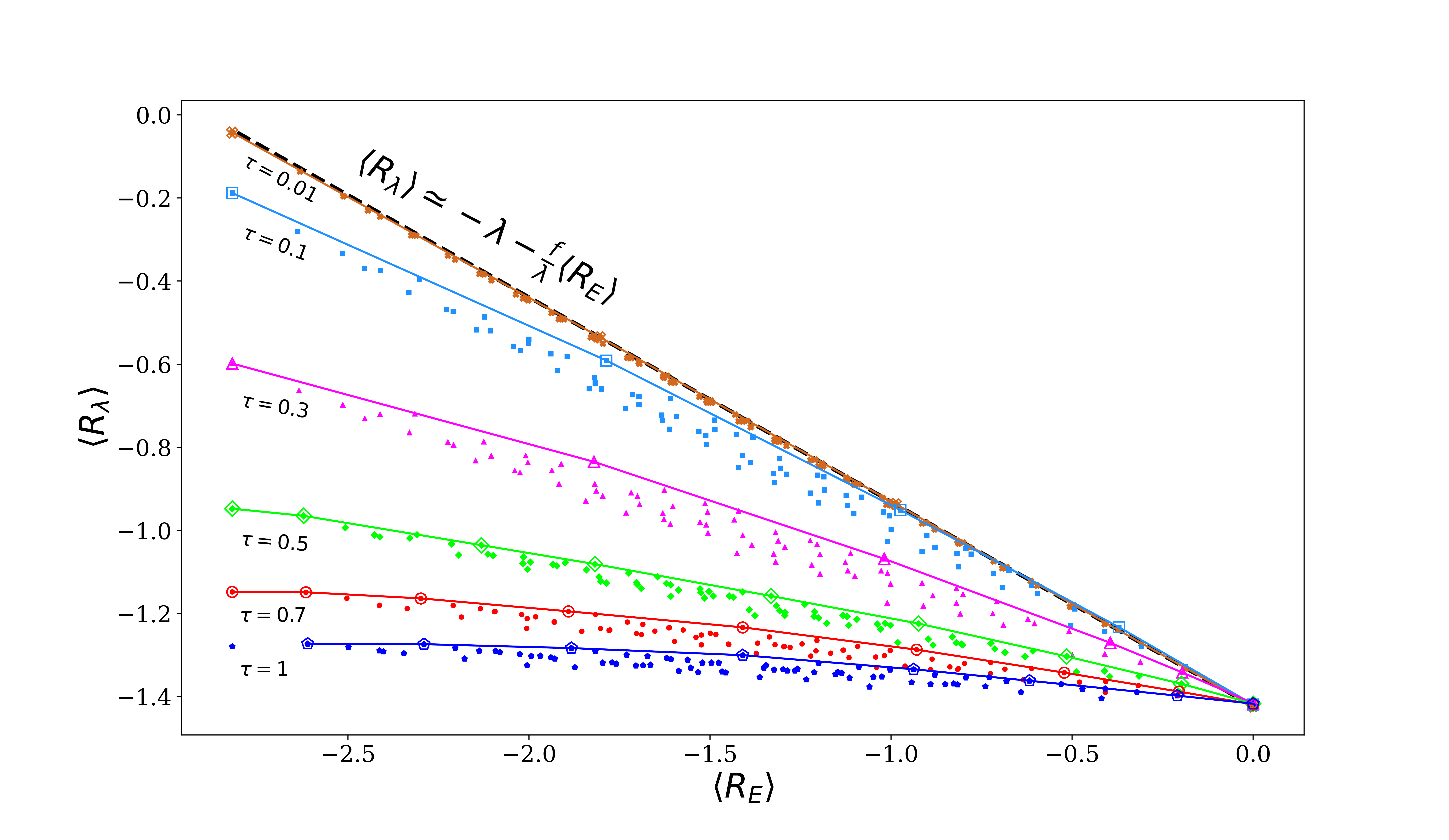} 
\caption{Pareto frontier for a \textit{reduced policies-space}, which is a lower bound to the true Pareto frontier:
  $\langle R_\lambda \rangle$ vs $\langle R_E \rangle$ obtained
  varying the decision time, $\tau$ for the 81 heuristic policies. The
  dashed black line indicates a linear dependence between $\langle
  R_\lambda \rangle$ and $\langle R_E \rangle$ reached for small
  $\tau$. We show with empty symbols the Pareto frontier that would
  result for each decision time by restricting the problem to study
  only the $81$ heuristic policies. It can be seen that as $\tau$
  increases, the frontier becomes concave.}
    \label{fig:Fig7}
\end{figure}

In Fig.\ref{fig:Fig7} we show the same plot of Fig.\ref{fig:Fig5} for
different values of $\tau$ and only considering the
heuristic policies, taking the Pareto dominating strategies in this
reduced policy space we obtain a lower bound to the true Pareto
frontier, which is enough for the following analysis.  Small decision values of $\tau$ correspond to frequent
measurements of the system and thus, as intuition
would suggest, to major adjustments of the
control variables that lead to high quality performances. In
particular, $\tau=0.01=dt$ (i.e. controlling at each time step, $dt$)
leads to a linear Pareto frontier, $\langle R_\lambda \rangle \propto
\langle R_E\rangle$, where all policies are equivalent, meaning that
they all live on the Pareto frontier: none (Pareto) dominates the
others.  In other terms, if the Lagrangian pair can continuously sense
the environment, does not really need to search for optimal policies.
Increasing the decision time values leads to concave frontiers where
the strategies play different roles until $\tau$ starts to be too
large with respect to $1/\lambda$ and swimming becomes ineffective in
controlling the separation growth and it only represents an energy
cost. For instance, for $\tau=1$ swimming does not help on minimizing
the separation and thus swimming or being passive is almost equivalent.

To better understand the linear behavior at $\tau\to 0$, we
can derive the following heuristic analysis. Since for the whole
duration of an episode we are in the linear regime of separation
(i.e. the agents see a differentiable velocity field), we know from
\eqref{eq:totalRlambda} that $-\langle R_\lambda \rangle$ is nothing
but the Lyapunov exponent of the controlled system, $\lambda_c$. The
actions of swimming along the joining direction 
introduce a clear contraction factor, so (if $\tau$ is sufficiently
small) we can estimate the controlled Lyapunov exponent for the
heuristic policies as follows:
\begin{equation}
    \lambda_c \simeq \frac{\lambda\,T_{max} - (2f\,T_2 - f\,T_1)}{T_{max}},
\end{equation}
where we have decomposed the total episode duration as
$T_{max}=T_0+T_1+T_2$, with $T_1$ being the average time in which only one agent is swimming, 
while $T_0$ and a $T_2$ is the average time in which both agent are passive or swimming, respectively.  For the
\textit{na\"{i}ve} baseline $T_2=T_{max},\,T_1=T_0=0$ and we can
estimate $\lambda_{\textit{na\"{i}ve}}\simeq \lambda-2f$ (if
$\tau\rightarrow 0$). This means that for $\lambda\simeq 1.4$ and
$f=0.7$ the \textit{na\"{i}ve} baseline should be very close to a
\textit{perfect control}, i.e. it can keep the distance constant.  On
the other hand, based on the same decomposition of the total time,
$\langle R_E\rangle$ can be approximated as
\begin{equation}
  \langle R_E\rangle =-\frac{\lambda}{T_{max}}(2T_2+T_1)\,,
\end{equation}  
which implies a linear dependence between $\langle R_\lambda \rangle$ and $\langle R_E\rangle$:
\begin{equation}\label{eq:linearity_Rlambda_Re}
    \langle R_\lambda \rangle \simeq-\lambda -\frac{f}{\lambda}\langle R_E\rangle  \,,
\end{equation}
which explains the linearity of the Pareto frontier for $\tau=dt$ in
Fig.\ref{fig:Fig7}.  When the above relation applies (i.e. for $\tau\to 0$)
we can write the total reward as 
\begin{equation}
    \langle R_{tot}\rangle \simeq -\lambda+ \langle R_E\rangle \Big{(}\beta -\frac{f}{\lambda}\Big{)}\,.
\end{equation}
It is now clear why all the policies lie on the Pareto frontier, and thus are equivalent. For $\beta < f/\lambda$ the task is to minimize $\langle R_E\rangle$ (remember that $\langle R_E\rangle$ is negative defined), that means controlling continuously the system; For $\beta > f/\lambda$ the goal is to maximize $\langle R_E\rangle$, which is  maximal (i.e. equal to $0$) when both agents are passive. For $\beta=\beta_c=f/\lambda$ all  policies perform the same, $\langle R_{tot}\rangle \simeq -\lambda$.

Clearly, the linearity of the frontier is due to the choice of the swimming penalization we adopted (Cfr. eq.~\eqref{eq:rE}), which is linear in the number of swimming agents. Different (nonlinear) choices would break the linearity but will not
invalidate the decomposition we used above. In this paper we are not
interested in exploring other definition of rewards and  we
have chosen the simplest definition, which is enough to highlight the 
non trivial role played by the discrete decision time. Indeed it is
due to such discreteness that the agents need to learn how to
exploit the flow in an intelligent way and where different policies
are not equivalent anymore in terms of performances.

\section{Discussions and conclusions}\label{sec5}

We have presented a multi-agent and multi-task problem set to
minimize, at the same time, the dispersion rate of a Lagrangian pair
dominated by Lagrangian Chaos, in a stochastic flow, and the energy
consumption due to the active control on the system.  We modeled the
agents with limited observation capabilities, 16 states inform them
on the longitudinal and transversal velocity gradients in a
discretized form, and with a set of 15
possible choices of action for the agent pair. Thus, the space of deterministic policies is very large and counts $15^{16}$ navigation strategies. Furthermore, the agents
could swim with a variable but limited velocity intensity, namely they
were not able to overcome the chaoticity properties of the system. To
solve this problem, we have developed a MORL approach based on the
combination of the simple Q-learning algorithm and the scalarization
technique; this enabled us to show a systematic investigation of the
problem studying the Pareto frontier. In this direction, we have shown
how controlling only at discrete decision times makes the problem
nontrivial.  Indeed, the larger the interval between decision times is the more control
variables performances become unpredictable or, at least, not easy to
guess a priori. In the limit of continuous control, $\tau\rightarrow 0$,
the problem reduces to a linear Pareto frontier, where all policies
are equivalent (i.e. they  all  live on the frontier), while
increasing $\tau$ the frontier becomes concave and the strategies play
different roles in minimizing the separation. Instead, for high
decision time, $\tau\gg 1/\lambda$, with $\lambda$ the Lagrangian
Lyapunov exponent of the system, swimming becomes ineffective to control the pair separation.

We stress that the MORL techniques here implemented is model-free, as
it requires only a few local and instantaneous information about the
underlying flow. It does not require an individual
optimization for each initialization, and it is generic.

Remarkably, we showed that, within our setup, RL is able to reach
solutions that are strongly different from a \textit{na\"{i}ve}
baseline and, in general, they are different from heuristic references
based on ``longitudinal'' actions only.  It would be important
to extend the present approach to the case of smart tracers able to control their separation within scales where 
velocity field is no more differentiable, i.e. in the inertial range
of turbulent flows. 

\section*{Acknowledgments}

This work was supported by the European Research Council (ERC) under the European Union’s Horizon 2020 research and innovation programme (Grant Agreement No. 882340).

\section*{Author contribution statement}
All authors conceived the research. CC performed all the numerical simulations and data analysis. All authors discussed the results. CC wrote the paper with revision and input from all the authors.

\section*{Data availability statement}
Data sharing not applicable to this article as no datasets were generated or analysed during the current study.

\begin{appendices}
\section{Q-learning implementation}\label{appendixA}
To solve the optimization problem we used the Q-learning algorithm \cite{sutton2018reinforcement} which is based on evaluating the action-value function,  $Q(s,a)$, that is the expected future cumulative reward given the agents are in state $s$ and take
action $a$.   The algorithm is expected to converge to the optimal policy by the following iterative trial-and-error protocol. 
At each decision time $t_j$, the agents pair measures its state $s_{t_j}$ and selects an
action $a_{t_j}$  using an $\epsilon$-greedy strategy, where 
$a_{t_j}(s_{t_j})=\arg\max_a\{Q(s_{t_j},a)\}$ with probability $1-\epsilon$ or $a_{t_j}$ is chosen randomly with probability $\epsilon$. Then, we let the dynamical system evolve for a time $\tau$, according to
\eqref{eq:dynamics}, keeping both control directions and velocity
intensity fixed. Afterwards, the agents receive a reward
$r_{tot}(t_{j+1})$ \eqref{eq:rtot} and the Q-matrix is updated as
\begin{equation}
\begin{split}
    Q(s_{t_j},a_{t_j}) \leftarrow \,& Q(s_{t_j},a_{t_j}) +
 \alpha[r_{tot}(t_{j+1})
    + \\& +\max_{a} Q(s_{t_{j+1}},a)-Q(s_{t_j},a_{t_j})]\,,
\end{split}
\end{equation}
where $\alpha$ is the learning rate. Updates are repeated up to the end of the episode  $t=T_{max}$,  when no reward is assigned.
The learning protocol is repeated restarting with another pair with the same initial distance in another flow position until  we reach a ``local'' optimum  given by the equation $Q^*(s_{t_j},a)= r_{tot}(t_{j+1})
 +\max_{a} Q^*(s_{t_{j+1}},a)$ and defined by the policy $$a(s)=\arg\max_a\{Q^*(s,a)\}.$$ 
 In order
to ease the convergence of the algorithm, the learning rate $\alpha$
is taken as a decreasing functions of the time spent in the
state-action pair, while the  exploration parameters decreases with
the time spent in the visited state. Thus if $n(s,a)$ is the number of
decision times in which the couple $(s,a)$ has been visited, and
\begin{equation}
n(s)=\sum_a n(s,a)/|\mathcal{A}|\,,
\end{equation}
$\epsilon$ and $\alpha$ are taken
as:
\begin{eqnarray}
  \alpha&=&5/[200^{1/\gamma}+\tau n(s,a)]^{\gamma} \label{eq:alpha}\\
    \epsilon&=&5/[200^{1/\gamma}+\tau n(s)]^{\gamma}\label{eq:eps}
  \end{eqnarray}
with $\gamma=4/5$, the numerical values of the constants have been
determined after some preliminary tests. As for the initialization of
the matrix $Q$ we have taken the same large (optimistic) value for all the state-action pairs.
\end{appendices}

\end{document}